\documentclass[12pt]{article}
\usepackage{epsf}
\usepackage{psfig}
\usepackage{epsfig}
\usepackage{graphicx}
\usepackage{amssymb}

\def\di{\displaystyle}

\def\tr{{\rm tr}}

\def\s0#1#2{\mbox{\small{$ \frac{#1}{#2} $}}}
\def\0#1#2{\frac{#1}{#2}}

\def\di{\displaystyle}

\def\bg{\begin{eqnarray}\begin{array}{rcl}\displaystyle}
\def\eg{\end{array} &\di    &\di   \end{eqnarray}}
\def\bm#1{\begin{eqnarray}\begin{array}{#1}\di}
\def\bmo#1{\begin{eqnarray*}\begin{array}{#1}\di}
\def\bml#1#2{\begin{eqnarray}\begin{array}{#1}\label{#2}\di}
\def\bgo{\begin{eqnarray*}\begin{array}{rcl}\displaystyle}
\def\ego{\end{array} &\di    &\di \nonumber  \end{eqnarray*}}

\def\btensor#1#2{\renew\left#1\begin{array}{#2}\di}
\def\brtensor#1#2#3{\ren#3\left#1\begin{array}{#2}}
\def\botensor#1#2{\renew\left#1\begin{array}{#2}}
\def\etensor#1{\end{array}\right#1}

\def\tr{{\rm tr}}

\def\id{1\!\mbox{l}}

\def\be{\begin{equation}}
\def\ee{\end{equation}}
\def\bea{\begin{eqnarray}}
\def\eea{\end{eqnarray}}

\def\h{\hbox{$\frac{1}{2}$}}

\date{\today}

\def\ren#1{\renewcommand{\arraystretch}{#1}}

\def\renew{\renewcommand{\arraystretch}{1}}
\ren{1.5}

\begin{document}

\begin{flushright}
TCDMATH 05-04\\
HD-THEP-05-10\\
\end{flushright}
\par
\vskip .1 truecm
\large \centerline{\bf Doubly Periodic Instanton Zero Modes}
\par
\vskip 0.5 truecm
\normalsize
\begin{center}
{\bf C.~Ford} ${}^{a}$ and
{\bf  J.~M.~Pawlowski} ${}^{b}$
\\
\vskip 0.5 truecm
${}^a$\it{
School of Mathematics\\
Trinity College, Dublin 2, Ireland.\\ 
{\small\sf ford@maths.tcd.ie}\\
}
                                            
\vskip 0.5 truecm
${}^b$\it{Institut f\"ur Theoretische Physik,\\
Universit\"at Heidelberg, Philosophenweg 16\\
D 69120 Heidelberg, 
 Germany. \\
{\small\sf J.Pawlowski@thphys.uni-heidelberg.de}
}

\vskip 0.5 true cm

\end{center}

\vskip .7 truecm\normalsize
\begin{abstract} 
  
  Fermionic zero modes associated with doubly periodic $SU(2)$
  instantons of unit charge are considered. In cases where the action
  density exhibits two `instanton cores' the zero mode peaks on one of
  four line-segments joining the two constituents.  Which of the four
  possibilities is realised depends on the fermionic boundary
  conditions; doubly periodic, doubly anti-periodic or mixed.

\end{abstract}

\baselineskip=20pt 

In this Letter we consider fermionic zero modes for the recently
discussed doubly periodic instantons \cite{Ford:2002pa,Ford:2003vi}.
Two complementary constituent descriptions of these objects were
provided; charge one instantons can be built out of two overlapping
instanton cores \it or \rm two static monopoles.  Explicit
computations show that in square tori the action density peaks at two
points in $T^2\times R^2$ in line with the core picture.  For
elongated tori with high aspect ratios the action density is
concentrated in two tubes which can be interpreted as the worldlines
of monopole constituents.  The basic properties of these monopoles
(separation and mass ratio) follow much the same pattern as found for
the monopole constituents of calorons
\cite{Nahm:1983sv,Lee:1998bb,Kraan:1998pm,Bruckmann}.  Here we compute
the zero mode density within a two-dimensional slice including the
constituent locations for various boundary conditions.  These results
are compared with the action density calculations reported in refs.
\cite{Ford:2002pa,Ford:2003vi}.  Of particular interest are the
localisation properties of the zero modes with respect to the
instanton core constituents and their evolution as the aspect ratio,
or temperature, is increased.

To begin we recall some basic definitions regarding gauge fields on
$T^2\times R^2$.  A doubly-periodic gauge potential is understood to
be an anti-hermitian potential defined throughout $R^4$ which is
periodic modulo gauge transformations $U_1$ and $U_2$ in two
directions
\begin{eqnarray}
A_{\mu}(x_0,x_1+L_1,x_2,x_3)&=&U_1\left(
A_{\mu}(x_0,x_1,x_2,x_3)+\partial_{\mu}\right)
U_1^{-1},\label{bcs}\\
A_{\mu}(x_0,x_1,x_2+L_2,x_3)&=&U_2\left(
A_{\mu}(x_0,x_1,x_2,x_3)+\partial_{\mu}\right)
U_2^{-1}.\nonumber\end{eqnarray}
In general the transition functions $U_1$ and $U_2$ are $x$-dependent.
However we will work in a gauge where they are constant commuting
group elements. Then $\hbox{tr } U_1$ and $\hbox{tr } U_2$ are the two
holonomies (assuming that $A_1$ and $A_2$ vanish at infinity). We
specialise to self-dual $SU(2)$ potentials in the one-instanton
sector.  The Weyl operator
\begin{equation}\label{eq:Weyl} 
D^{\dagger}(A)=-\sigma_{\mu}^{\dagger}\left(\partial_{\mu}+A_{\mu}\right)\,
\end{equation} 
with $\sigma_\mu^\dagger=(1,-i\tau_1,-i\tau_2,-i\tau_3)$ and $\tau_i$
are Pauli matrices, is expected to have a single fermionic zero mode.
To specify the periodicity properties of the fermionic zero mode two
phases are required
\begin{eqnarray} 
\Psi(x_0,x_1+L_1,x_2,x_3;z)&=&e^{iz_1 L_1}U_1\Psi(x_0,x_1,x_2,x_3;z)
\label{fermionbcs}\\
\Psi(x_0,x_1,x_2+L_2,x_3;z)&=&e^{iz_2 L_2}U_2\Psi(x_0,x_1,x_2,x_3;z).
\nonumber
\end{eqnarray}
To make contact with the Nahm formalism the phases are parametrised by
dimensionful coordinates $z_1$ and $z_2$ rather than angles; these
have the interpretation as coordinates of the dual torus, $\tilde
T^2$, since the replacements $z_1\rightarrow z_1+2\pi/L_1$
$z_2\rightarrow z_2+2\pi/L_2$ leave the boundary conditions unchanged.
Such general boundary conditions have also been studied in a lattice
context \cite{Gattringer:2004va}.  The choice $z_1=z_2=0$ leads to
periodic fermions while $z_1=\pi/L_1$, $z_2=\pi/L_2$ provides
`physical' anti-periodic solutions.  Another interesting case is when
$z_1$ and $z_2$ are correlated with the two holonomies.

The transition functions can be parametrised as follows
\begin{equation}
U_1=e^{-i\omega_1L_1\tau_3},~~~~
U_2=e^{-i\omega_2L_2\tau_3},
\end{equation}
where $\omega=(\omega_1,\omega_2)$, like $z=(z_1,z_2)$, can be
considered an element of the dual torus. Note that there is no
charge-one instanton solution for trivial holonomy $\omega=0$.  Like
$R^4$ instantons and calorons the $k=1$ solution has a scale
parameter, $\lambda$, which can be thought of as the instanton size.
The scale parameter fixes another property of the instanton namely its
\it flux \rm $\kappa$; asymptotically the instanton has the form
\begin{equation}
A_{\mu}(x)\sim a_{\mu}(x_0,x_3)\tau_3,\end{equation}
where $a_\mu$ is a $U(1)$ self-dual potential in $R^2$.
The flux is defined through
\begin{equation}
\kappa=
\lim_{R\rightarrow\infty}
{i\over{2\pi}}\int_{C(R)}\left(a_0~dx_0+a_3~dx_3\right),
\end{equation}
where $C(R)$ is a circle of radius $R$ in the $x_0-x_3$ plane.  The
sign of $\kappa$ is ambiguous since the signs of the $a_\mu$ can be
flipped via a constant gauge transformation (a Weyl reflection).
Moreover, we may assume that $\kappa$ lies between $0$ and $1$ since
$\kappa$ can be changed by an integer amount via a smooth gauge
transformation.  The asymptotic flux can also have non-zero components
in the compact $x_1$ and $x_2$ directions, i.e.  $a_1$ and $a_2$ need
not be zero.  When $a_1=a_2=0$ the instanton has a radial symmetry;
the action density depends on $x_1$, $x_2$ and $r=\sqrt{x_0^2+x_3^2}$
only.  In this case the action density decays exponentially and there
is a simple relation between the scale parameter $\lambda$ and the
flux
\begin{equation}
\kappa=\frac{\pi\lambda^2}{L_1L_2}.
\end{equation}
These special radial solutions have seven parameters; the flux
$\kappa$ (or equivalently the size $\lambda$) the two holonomies and
four translations in $T^2\times R^2$.  Together the
$\kappa\rightarrow-\kappa$ ambiguity and the $\kappa\equiv \kappa+1$
equivalence imply that the fluxes $\kappa$ and $1-\kappa$ are
physically indistinguishable.  This gives two possible instanton sizes
$\lambda_1=\sqrt{\kappa L_1 L_2/\pi}$ and $\lambda_2=\sqrt{(1-\kappa)
  L_1 L_2/\pi}$.  In \cite{Ford:2002pa} it was argued that the
instanton possesses two \sl instanton core \rm constituents with sizes
$\lambda_1$ and $\lambda_2$.  Taking $x_\mu=0$ as the position of the
first core the second is centred at $x_1=L_1 L_2 \omega_2/\pi$,
$x_2=-L_1 L_2 \omega_1/\pi$, $x_0=x_3=0$, i.e. the core separation is
fixed by the holonomies.  For square tori ($L_1=L_2$) explicit
calculations of the action density clearly show two instanton-like
peaks at the expected locations.

In the special case $\kappa=\frac{1}{2}$ the two cores are identical.
Here the flux can be interpreted as a center vortex.  These solutions
are decompactified four torus instantons of unit charge (the
$L_0,L_3\rightarrow\infty $ limit of an $SU(2)$ instanton on $T^4$
with periods $L_0$, $L_1$, $L_2$ and $L_3$).  The center vortex is a
remnant of a torus twist, $Z_{03}=-\id$, see ref. \cite{'tHooft:1981sz}.
Because of the exponential decay, the $\kappa=\frac{1}{2}$ solutions
are expected to approximate four torus instantons with large but
finite $L_0$ and $L_3$ extremely well (see also
\cite{Gonzalez-Arroyo:1998ez}).  These doubly periodic instantons can
be seen as the opposite extreme to 't Hooft's constant curvature
solutions which exist when $L_1L_2=2L_0L_3$.  An analytic
interpolation between these two regimes is still lacking (see however
\cite{GarciaPerez:2000yt}).  In the absence of analytic solutions a
constituent description (in terms of cores, monopoles or otherwise) as
well as information concerning the moduli-spaces and their metrics
would represent a considerable advance.

If one period is much larger than the other, say $L_1>>L_2$, the core
picture fails; the action density is concentrated around two monopole
worldlines.  These monopole constituents follow a similar pattern to
that observed for charge one calorons; $\omega_2$ determines the mass
ratio of the two monopoles and their spatial separation is
$\pi\lambda^2/L_2=\kappa L_1$.  The caloron zero mode
\cite{GarciaPerez:1999ux} localises to one of the monopole
constituents according to the value of $z$.  As $z$ passes through a
critical value (where $z$ is correlated with the holonomy) the zero
mode switches its support to the other monopole.  If $z$ is exactly at
a critical value the zero mode peaks at both monopole locations.
Furthermore these delocalised zero modes do not decay exponentially
(the decay is sufficiently fast to give a normalisable solution).  In
the doubly periodic case we have to distinguish between the core
($L_1\approx L_2$) and monopole ($L_1>>L_2$ or $L_2>>L_1$) regimes.
If $L_1>>L_2$ the fermionic zero mode is expected to be caloron-like
in that it will localise to one monopole for $-\omega_2<z_2<\omega_2$
and the other for $\omega_2<z_2<-\omega_2+2\pi/L_2$.  If
$z_2=\pm\omega_2$ the zero mode will see both monopoles (assuming
$z_1\neq\pm \omega_1$, since as we shall argue
$(z_1,z_2)=\pm(\omega_1,\omega_2)$ are \it very \rm special cases).
What is less obvious is how the zero mode behaves in the core regime.
We have computed the zero mode density $\Psi^\dagger(x;z)\Psi(x;z)$
within the two-dimensional slice $x_0=x_3=0$ for various choices of
$\kappa$, $\omega$ and $z$.  When $L_1=L_2$, the zero mode localises
to one of four lines joining the cores.

\begin{figure}[h]
\begin{picture}(0,155)(-250,0)
\put(-250,0){\epsfig{file=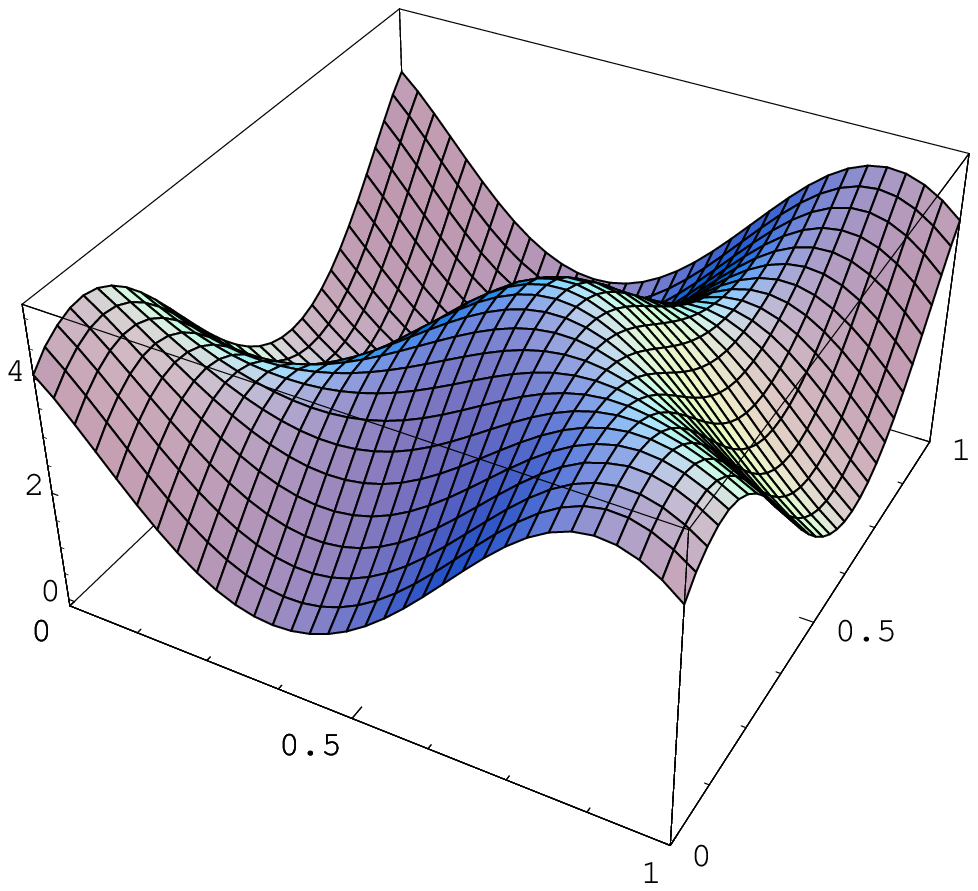,width=.4\hsize}}
\put(-250,140){$\Psi^\dagger\Psi$} 
\put(0,140){$-\s012 \tr\, F^2$} 
\put(-202,2){$\bf  x_2$}  
\put(-75,26){ $\bf x_1$}
\put(0,0){\epsfig{file=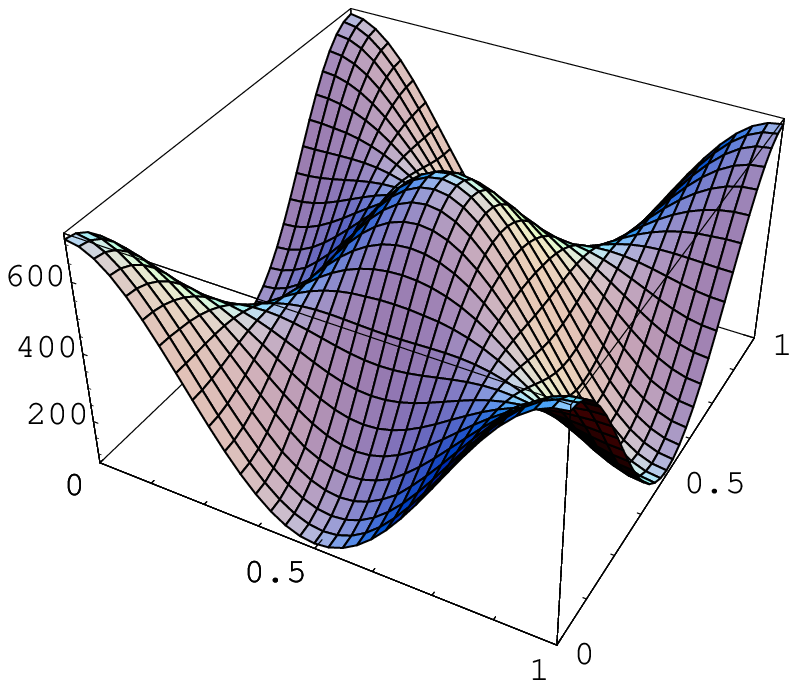,width=.4\hsize}}
\put(48,8){$\bf  x_2$}  
\put(175,39){ $\bf x_1$}
\end{picture}
\caption{zero mode density $\Psi^\dagger\Psi$ and action density 
$-\s012 \tr F^2$ 
for $x_\perp=0$,  
$\kappa=\s012$, $\omega=\s0\pi2(1,1)$ and $z=\pi(1,1)$ (anti-periodic).}
\label{cores}
\end{figure}

Consider the case $L_1=L_2=1$, $\kappa=\frac{1}{2}$,
$\omega_1=\omega_2=\frac{1}{2}\pi$.  Here the two (equal sized) cores
are particularly well resolved in the action density. They are located
at the origin $(x_1,x_2)=(0,0)$, and in the centre of the torus
$(x_1,x_2)=(\s012,\s012)$.  For $z_1=z_2=\pi$, the physical
anti-periodic case, the zero mode is not localised on a single core
but smeared around a line-segment joining the core at the corner
$(x_1,x_2)=(0,1)$ and core in the middle of the torus
$(x_1,x_2)=(\s012,\s012)$, see the left plot in figure~\ref{cores}.

\begin{figure}[h]
\begin{picture}(0,170)(-250,0)
\put(-250,0){\epsfig{file=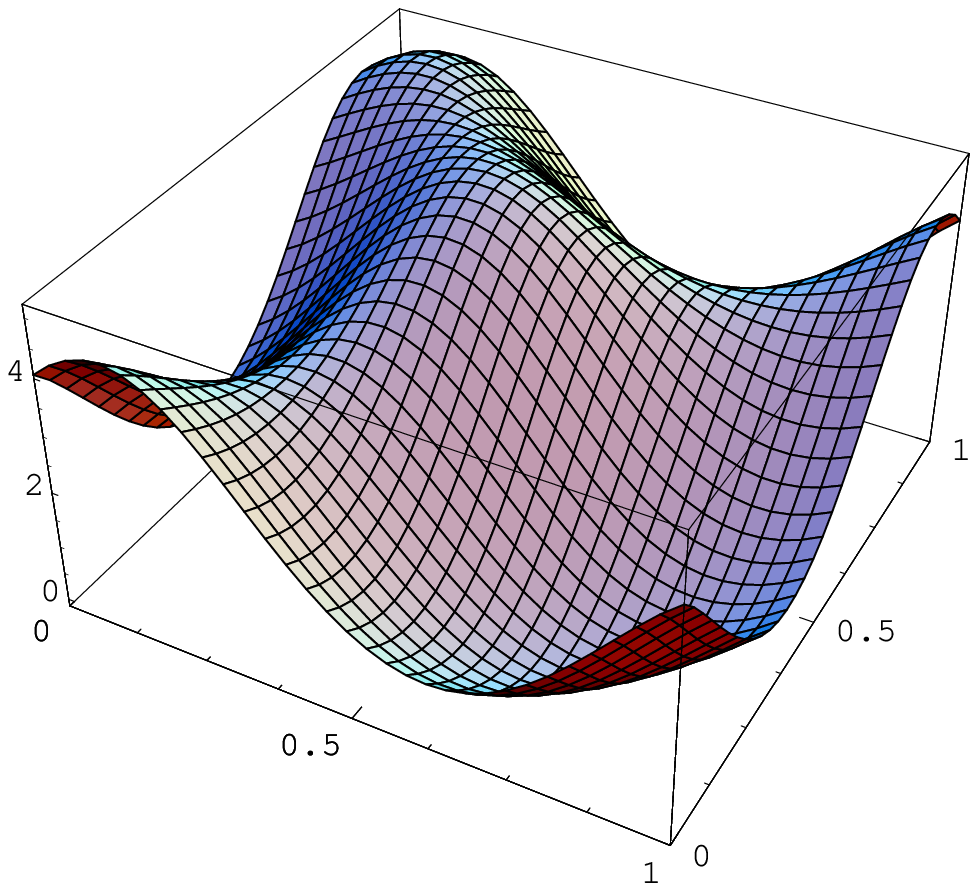,width=.4\hsize}}
\put(-250,122){$\Psi^\dagger\Psi$} 
\put(-250,155){$z=0$} 
\put(-202,2){$\bf  x_2$}  
\put(-65,46){ $\bf x_1$}
\put(0,0){\epsfig{file=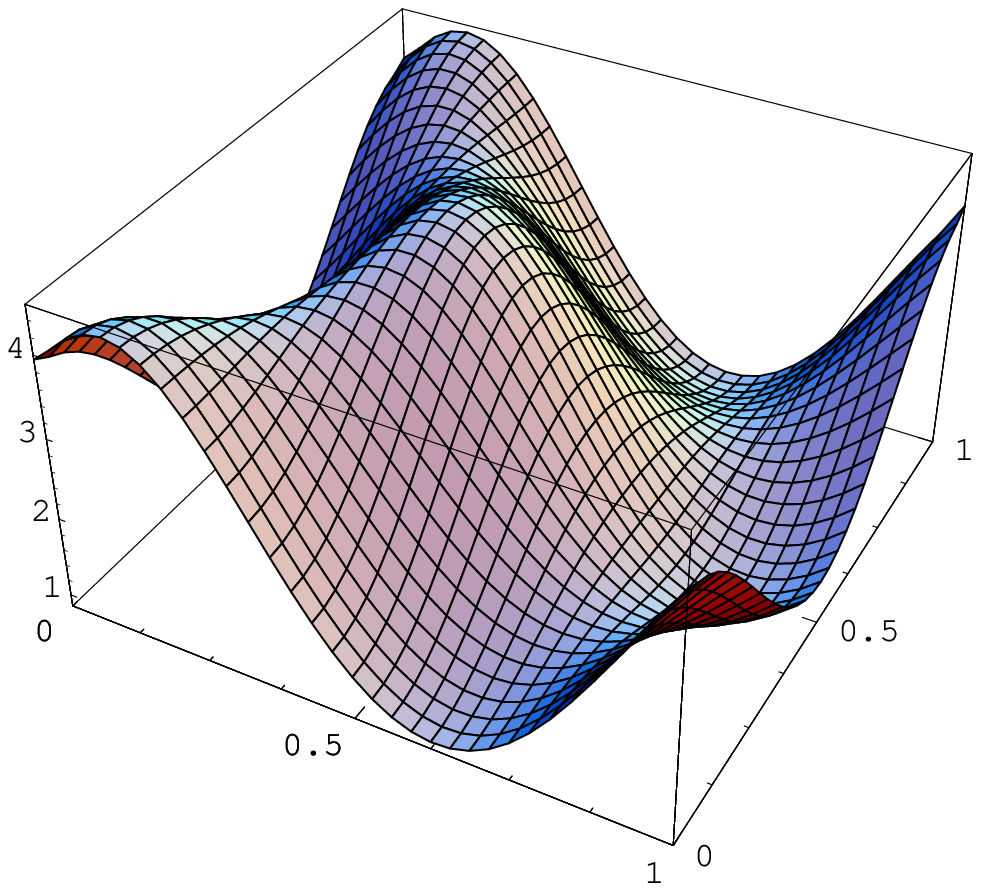,width=.4\hsize}}
\put(0,122){$\Psi^\dagger\Psi$} 
\put(0,155){$z=(0,\s0{\pi}{2})$} 
\put(48,-2){$\bf  x_2$}  
\put(175,39){ $\bf x_1$}
\end{picture}
\caption{zero mode density $\Psi^\dagger\Psi$ for $x_\perp=0$,  
$\kappa=\s012$  and $z=(0,0)$ 
(periodic) and  
$z=(0,\s0\pi2)$ (periodic in $x_1$ and anti-periodic in $2 x_2$).}
\label{zdirections}
\end{figure}

Taking instead $z=0$, the periodic case, yields a similar
line-localisation but with a different pairing arrangement; it
stretches between the core in the middle and that at
$(x_1,x_2)=(1,0)$, see the left plot in figure~\ref{zdirections}.

The zero modes densities in figures~\ref{cores} and \ref{zdirections}
(left plots) can be mapped into each other by a 180 degree rotation.
The other two pairing arrangements are given by $z=(0,\pi)$, where the
zero mode stretches between the cores at $(x_1,x_2)=(0,0)$ and
$(x_1,x_2)=(\s012,\s012)$ and $z=(\pi,0)$, where the zero mode
stretches between the cores at $(x_1,x_2)=(1,1)$
and $(x_1,x_2)=(\s012,\s012)$.  The transition
between two of the four possible arrangements is a smooth one.  For
example, taking $z=(0,\pi/2)$ generates a superposition of two of the
four generic pairings, namely between $z=(0,\pi)$ and $z=(0,0)$, see
the right plot in figure~\ref{zdirections}.

\begin{figure}[h]
\begin{picture}(0,431)(-250,0)

\put(-250,300){\epsfig{file=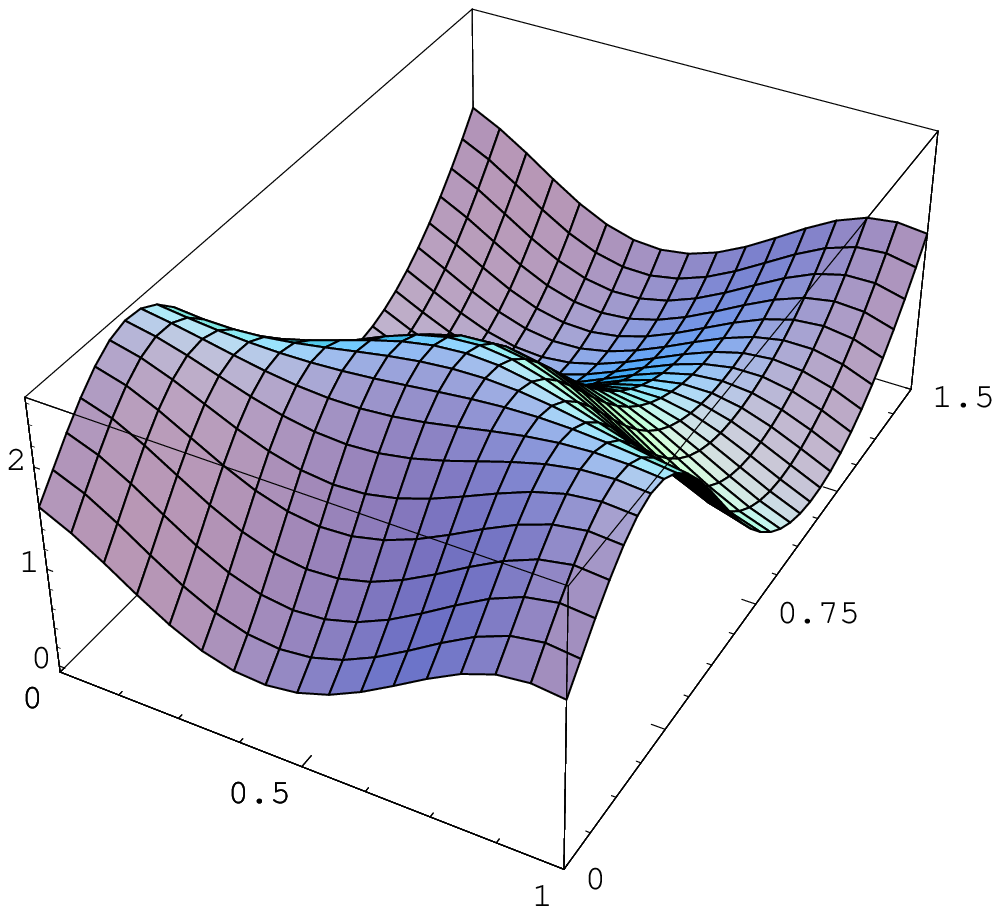,width=.3\hsize}}
\put(-250,395){$\Psi^\dagger\Psi$} 
\put(-20,395){$-\s012 \tr F^2$} 
\put(-80,420){$\bf a=\frac{3}{2}$} 
\put(-210,300){$\bf  x_2$}  
\put(-125,326){ $\bf x_1$}
\put(0,300){\epsfig{file=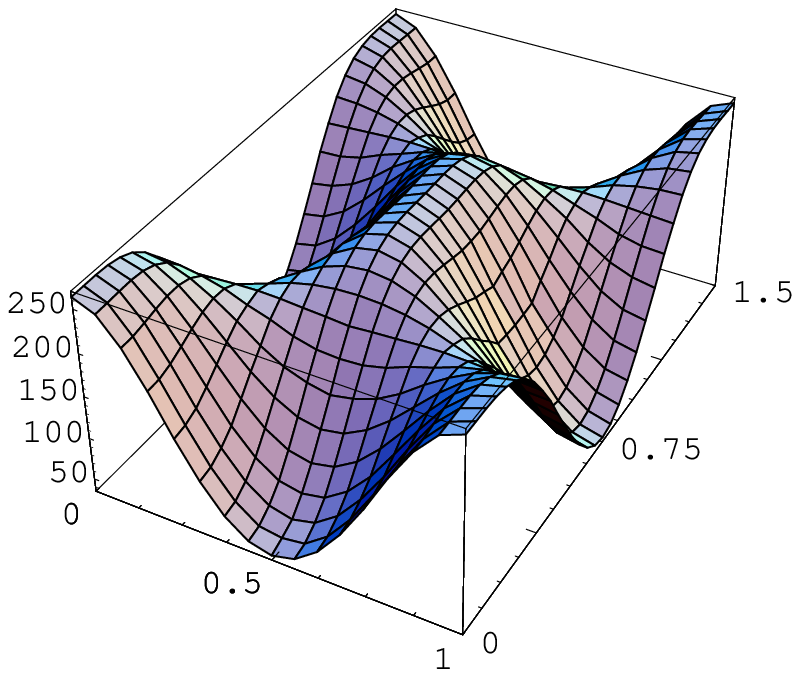,width=.3\hsize}} 
\put(38,302){$\bf  x_2$}  
\put(115,335){ $\bf x_1$}

\put(-250,150){\epsfig{file=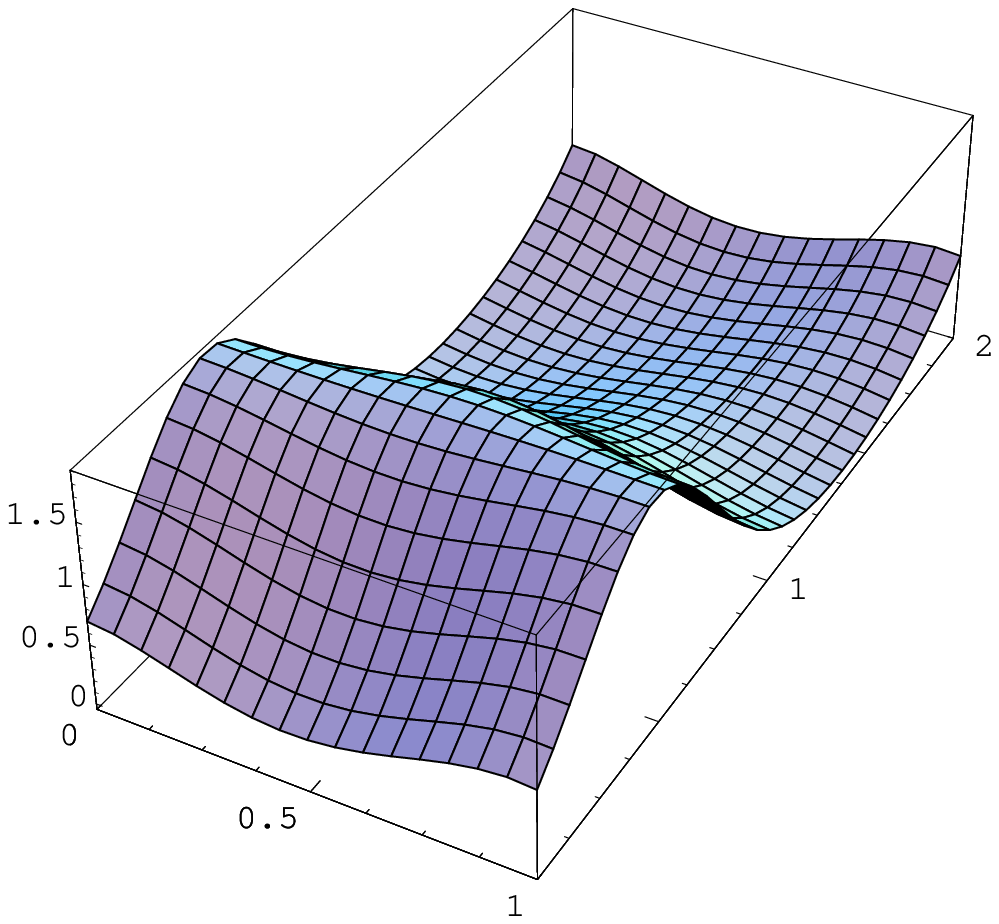,width=.3\hsize}}
\put(-80,270){$\bf a=2$} 
\put(0,150){\epsfig{file=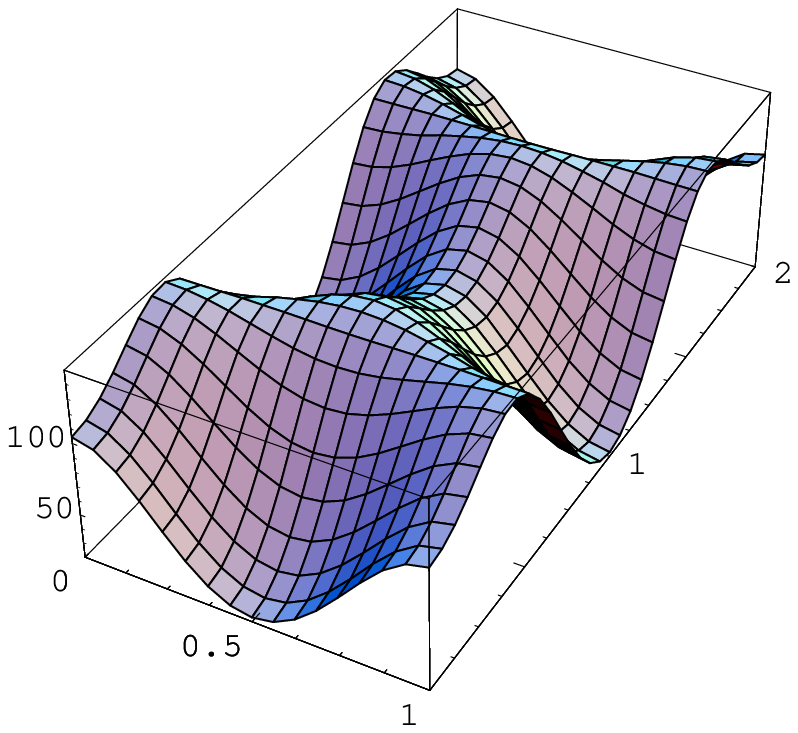,width=.3\hsize}}

\put(-250,0){\epsfig{file=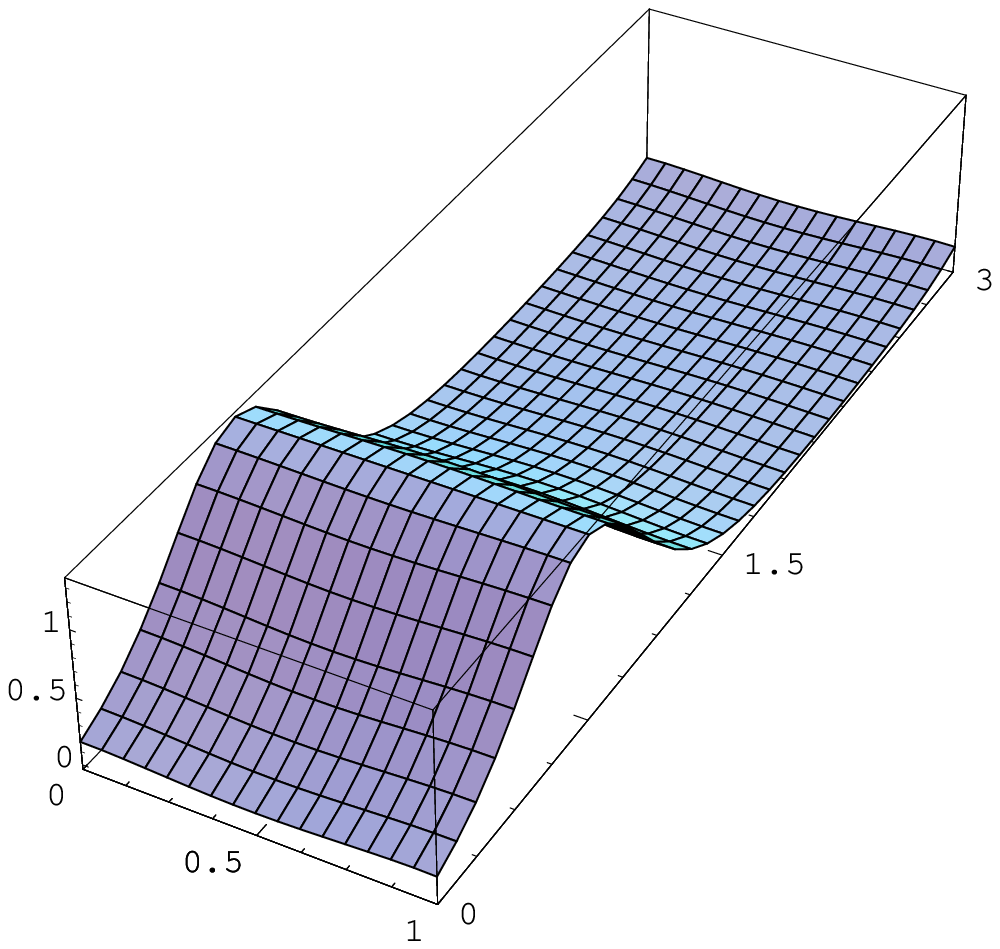,width=.3\hsize}}
\put(-80,130){$\bf a=3$} 
\put(0,0){\epsfig{file=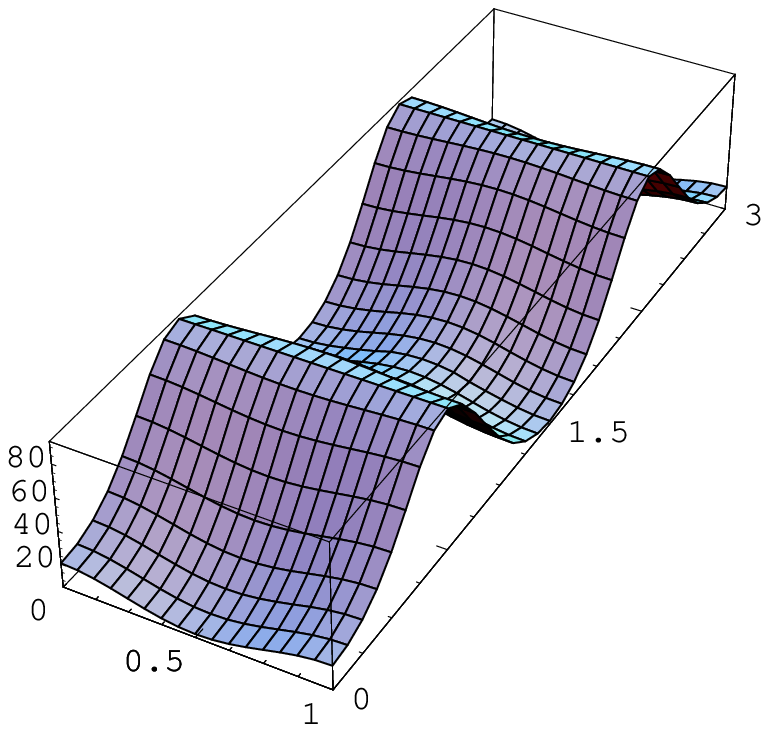,width=.3\hsize}}
\end{picture}
\caption{zero mode density $\Psi^\dagger\Psi$ 
for $x_\perp=0$
$\kappa=\s012$  and $\omega=\s0\pi2(a^{-1},1)$ and  
$a=\s032,2,3$ (anti-periodic)}
\label{history}
\end{figure}

The transition from the above situation to the finite temperature case
is accessed by increasing $L_1$ (or $L_2$) starting from $L_1=L_2$.
This was done in \cite{Ford:2003vi} for the action density showing the
crossover from instanton cores to monopole constituents. In
figure~\ref{history} we show the corresponding zero mode densities for
zero modes with anti-periodic boundary conditions,
 and aspect ratios $a=L_1/L_2=\s032, 2,3$.

The equal length case $a=1$ is already shown in figure~\ref{cores}.
If $a>1$ the zero mode peaks at one of the monopole worldlines. In
fact the monopole structure emerges in the zero modes \it before \rm
it is visible in the action density; the $a=\frac{3}{2}$ zero mode
already has some resemblance to the large $a$ monopole like regime
whereas (stretched) cores are still visible in the action density.  In
figure 4 we show zero mode and action densities for the case
$\kappa=\frac{5}{8}$. In the $a=1$ core regime the zero mode density
peaks at the smaller core but one still can see a preferred line
joining the smaller core at the centre to the larger core at the
corner $(x_1,x_2)=(0,1)$.  Here the evolution to the monopole regime
is slower due to the presence of a smaller core.  But as with the
$\kappa=\frac{1}{2}$ case the monopole structure appears in the zero
mode density before it can be seen in the action density.

\begin{figure}[h]
\begin{picture}(0,300)(-250,0)

\put(-250,150){\epsfig{file=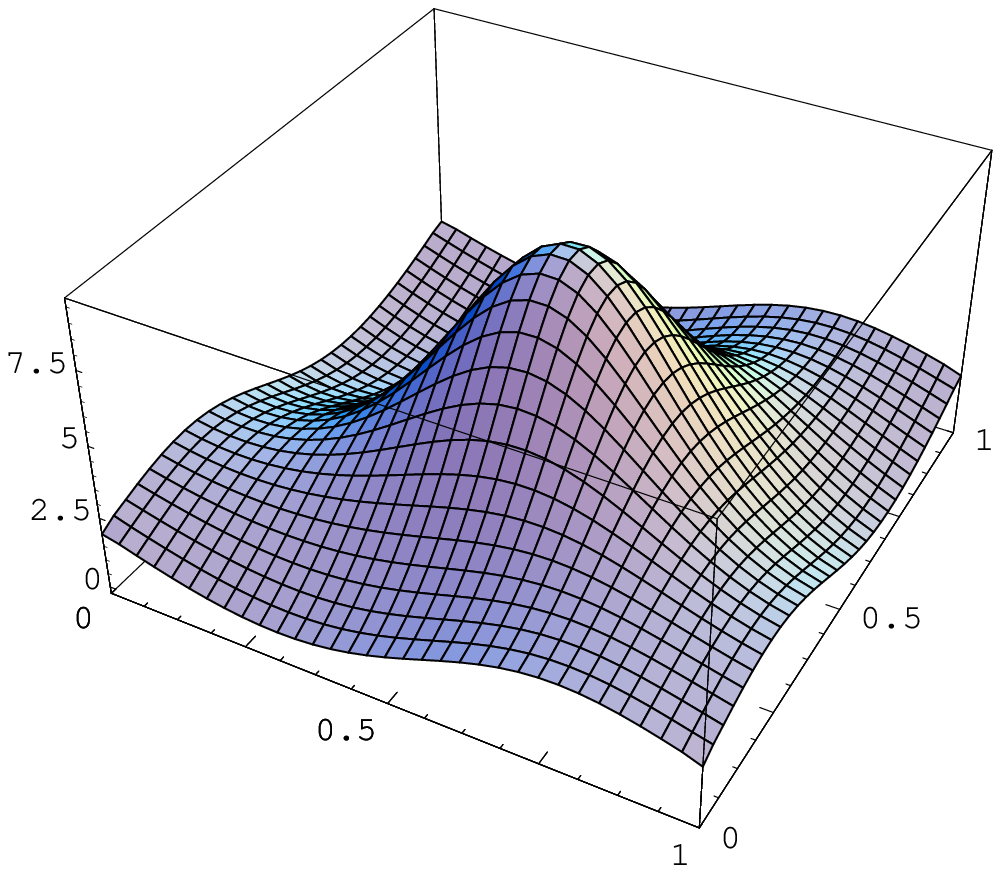,width=.3\hsize}}
\put(-250,245){$\Psi^\dagger\Psi$} 
\put(-20,255){$-\s012 \tr F^2$} 
\put(-80,270){$\bf a=1$} 
\put(-210,150){$\bf  x_2$}  
\put(-125,176){ $\bf x_1$}
\put(0,150){\epsfig{file=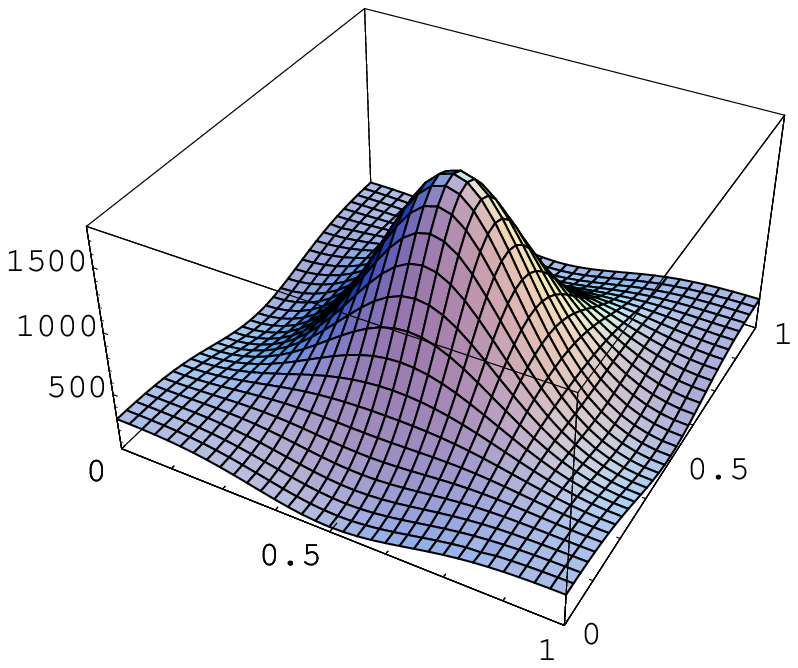,width=.3\hsize}} 
\put(43,152){$\bf  x_2$}  
\put(125,176){ $\bf x_1$}

\put(-250,0){\epsfig{file=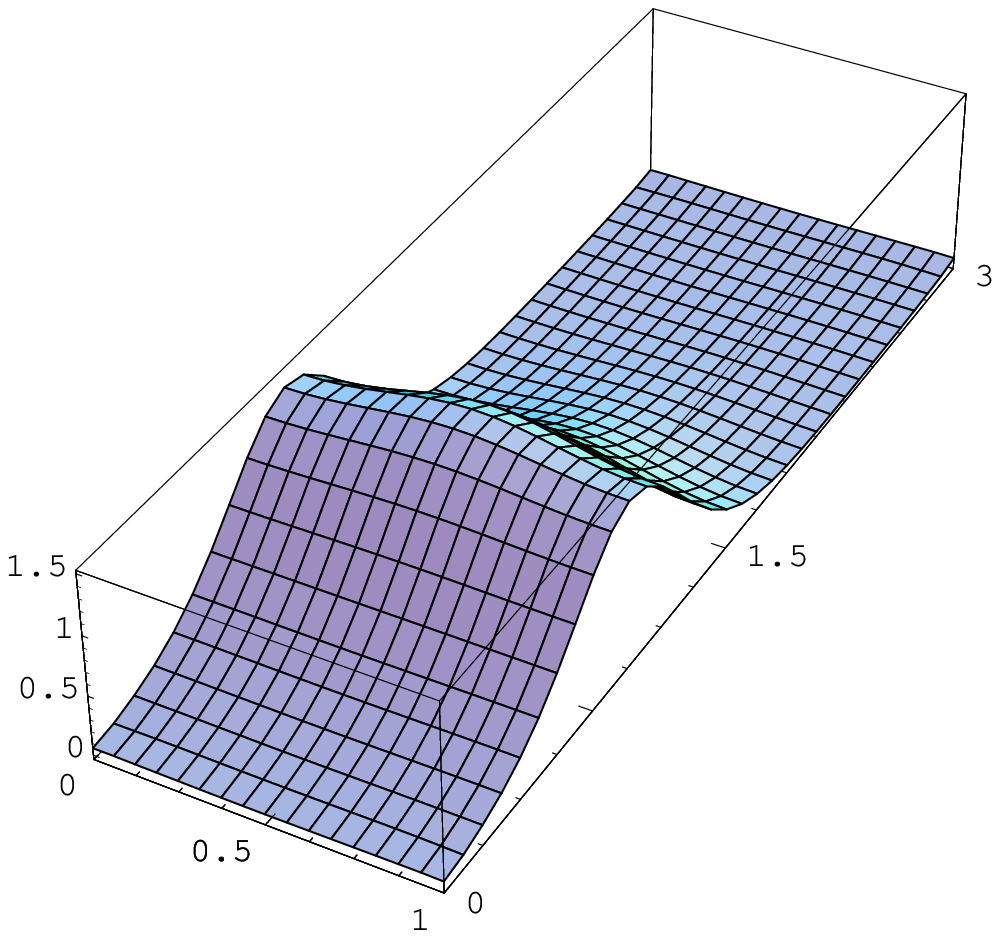,width=.3\hsize}}
\put(-80,120){$\bf a=3$} 
\put(0,0){\epsfig{file=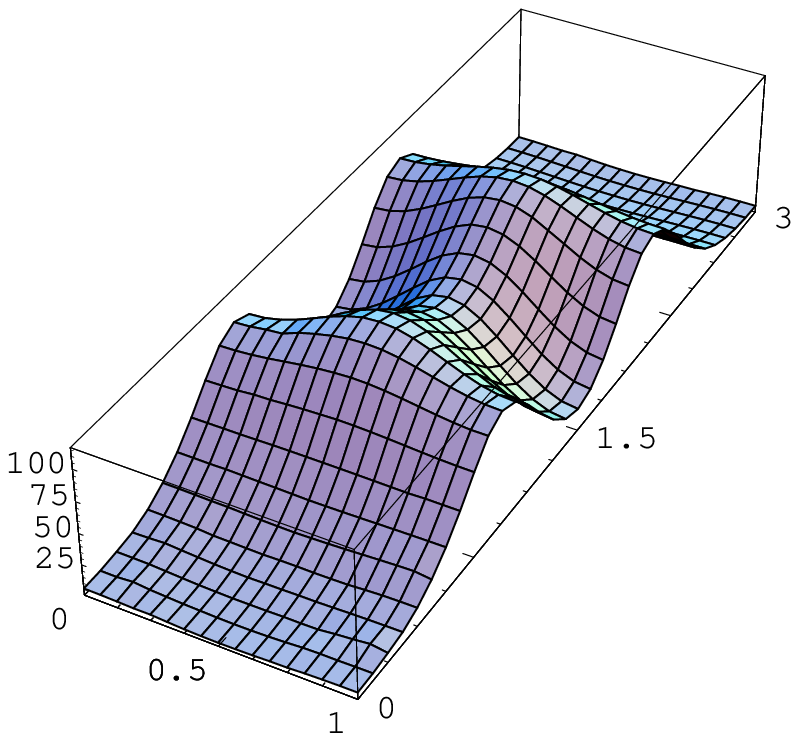,width=.3\hsize}}
\end{picture}
\caption{zero mode density $\Psi^\dagger\Psi$ 
  for $x_\perp=0$, $\kappa=\s058$ and $\omega=\s0\pi2(a^{-1},1)$ and
  $a=1,3$ (anti-periodic)}
\label{history58}
\end{figure}

A further interesting case concerns the limit where $z$ approaches
$\pm\omega$.  If $\kappa\neq\frac{1}{2}$ the zero mode does localise
to a single core as $z$ \it approaches \rm $\omega$ or $-\omega$.
However \it at \rm $z=\omega$ the zero mode is not normalisable.  If
$\Psi(x,z)$ is normalised then for a fixed $x$ it tends to zero as $z$
tends to $\omega$.  When $\kappa=\frac{1}{2}$ the zero mode also
becomes non-normalisable as $z$ approaches $\omega$ but during the
approach the pair structure survives, i.e. however close $z$ is to
$\omega$ the zero mode will not be localised to a single core.  When
$z=\omega$ the equation $D^\dagger \Psi=0$ has two solutions
$\Psi^I(x)$ and $\Psi^{II}(x)$ which are supported at the first and
second core, respectively.  These `zero modes' are smooth but not
normalisable.  Like the $z=\omega$ caloron zero mode they decay
algebraically - but not fast enough to be normalisable.  As $z$
approaches $\omega$ the zero mode has the form
\begin{equation}
\Psi(x;z)\sim
|z-\omega|^{\kappa}\Psi^I(x)+|z-\omega|^{-\kappa}
(z_1+iz_2-\omega_1-i\omega_2)\Psi^{II}(x).
\end{equation}
As $z$ approaches $\omega$ the $\Psi^{II}$ contribution is suppressed
if $\kappa<\frac{1}{2}$ and for $\kappa>\frac{1}{2}$ the first term is
suppressed. Accordingly, the zero mode is localised at the smaller
core.  If $\kappa=\frac{1}{2}$ neither core is favoured and the pair
localisation persists. Note that the zero mode density depends on the
phase of $z_1+iz_2-\omega_1-i\omega_2$ so that the form of
$\Psi^{\dagger}\Psi$ for $z\sim\omega$ depends on the direction of
approach to $\omega$.  This is why all four pairings can be seen in
the $z\rightarrow\omega$ limit when $\kappa=\frac{1}{2}$.

The plots presented in this Letter were generated using an explicit
formula for $\Psi(x;z)$ valid in the two-dimensional slice
$x_0=x_3=0$.  To conclude we outline the derivation of this formula.
The construction hinges on the fact that the $SU(2)$ gauge potential,
$A_{\mu}(x)$, has a simple abelian Nahm transform, $\hat A(z)$, with
components \cite{Ford:2003vi}
\begin{equation}
\hat A_1(z)=-i\partial_{z_2}\phi(z),~~~
\hat A_2(z)=i\partial_{z_1}\phi(z),~~~\hat A_0(z)=\hat A_3(z)=0,
\end{equation}
where $\phi(z)$ is doubly-periodic and harmonic except at two flux
singularities in $\tilde T^2$.  This is the form of the Nahm potential
associated with radially symmetric one instantons on $T^2\times R^2$.
In the non-radial case, which we do not consider here, $\hat A_0$ and
$\hat A_3$ are non-zero (see also \cite{Jardim:1999dx}).  The
instanton can be expressed as a Nahm transform of $\hat A(z)$:
\begin{equation}
A_{\mu}^{pq}(x)=\int_{\tilde T^2}~d^2z~\psi^p{}^\dagger(z;x)
\frac{\partial}{\partial x^\mu}\psi^q(z;x),\end{equation}
where $\psi^p(z;x)$ ($p=1,2$) are orthonormal and periodic (with
respect to $z_1\rightarrow z_1+2\pi/L_1$ and $z_2\rightarrow
z_2+2\pi/L_2$) zero modes of the Weyl operator
\begin{equation}
D_x^\dagger(\hat A)=-\sigma_\mu^\dagger D_x^\mu(\hat A),
\end{equation}
where $D_x^\mu(\hat A)=\partial/\partial z_\mu+\hat A_\mu(z)-ix_\mu$
for $\mu=1,2$ and $D_x^\mu(\hat A)=\hat A_\mu(z)-ix_\mu$ for
$\mu=0,3$.  The Nahm zero modes can be written in the form
\begin{equation}
\psi^1(z;x)=D_x(\hat A)\left(\begin{array}{cr}
\varphi^{(1)}(z;x)\\0\end{array}\right),
\qquad \qquad 
\psi^2(z;x)=D_x(\hat A)
\left(\begin{array}{cr}
0\\
\varphi^{(2)}(z;x)
\end{array}\right)
,\label{nahmmodes}
\end{equation}
where the $\varphi^{(p)}$ are specific singular solutions of the
$\tilde T^2$ Laplace equation
\begin{equation}
\left(D_x^\mu(\hat A)\right)^2
\varphi(z;x)=0.
\end{equation}
The fermionic zero mode $\Psi(x;z)$ can be written in a similar
fashion to the Nahm zero modes
\begin{equation}
\Psi(x;z)=D( A)\left(\begin{array}{cr}
\Phi^{(1)}(x;z)\\0\end{array}\right)
=D( A)
\left(\begin{array}{cr}
0\\
\Phi^{(2)}(x;z)
\end{array}\right)
,\label{zmode}
\end{equation}
where $\Phi^{(1)}$ and $\Phi^{(2)}$ are specific singular solutions of
the $T^2\times R^2$ Laplace equation
\begin{equation}
\left(\frac{\partial}{\partial x_\mu}+A_\mu(x)\right)^2\Phi(x;z)=0.
\end{equation}
In fact, \it one \rm of the components of $\Phi^{(p)}$ is
$\varphi^{(p)}$
\begin{equation}
\Phi^{(1)}(x;z)=\left(\begin{array}{cr}
\tilde \varphi^{(1)}(z;x)\\ \varphi^{(1)}(z;x)\end{array}\right),
\qquad \qquad \Phi^{(2)}(x;z)=
\left(\begin{array}{cr}
\varphi^{(2)}(z;x)\\
\tilde\varphi^{(2)}(z;x)
\end{array}\right).\end{equation}
The other components, $\tilde\varphi^{(1)}$ and $\tilde\varphi^{(2)}$,
can be obtained from the requirement that $\Phi^{(1)}$ and
$\Phi^{(2)}$ generate the same zero mode, i.e. equation (\ref{zmode}).
This requirement amounts to four first order PDEs for
$\tilde\varphi^{(1)}$ and $\tilde\varphi^{(2)}$.  The integrability
condition for these equations can be expressed as another Laplace-type
equation
\begin{equation}
\left(\frac{\partial}{\partial x_\mu}+A_\mu^B(x)\right)^2
\left(\begin{array}{cr}
\varphi^{(1)}(z;x)\\ \varphi^{(2)}(z;x)\end{array}\right)=0,
\end{equation}
where $A_\mu^B(x)$ is an $SU(1,1)$ self-dual potential which is
related to $A_\mu(x)$ by a simple B\"acklund-type transformation; more
details of this structure will be given elsewhere.

The Nahm zero modes (\ref{nahmmodes}) lead to an instanton potential
\begin{eqnarray}
A_{x_{||}}&=&
-\frac{\tau_3}{2}\partial_{x_{||}}\log\rho
-2\pi i(\tau_1-i\tau_2)\kappa \rho \partial_{\bar x_\perp}
\frac{\nu^*}{\rho}
\nonumber\\
A_{x_{\perp}}&=&-
\frac{\tau_3}{2}\partial_{x_{\perp}}\log\rho
+2\pi i(\tau_1-i\tau_2)\kappa \rho \partial_{\bar x_{||}}
\frac{\nu^*}{\rho},\label{gaugepotential}
\end{eqnarray}
and $A_{\bar x_{||}}=-A_{ x_{||}}^\dagger$, $A_{\bar x_{\perp}}=-A_{
  x_{\perp}}^\dagger$, where $\rho(x)$ is real and periodic and
$\nu(x)$ is complex and periodic up to a constant phase.  Here we have
used two sets of complex coordinates for $T^2\times R^2$; in the
compact directions $x_{||}=x_1+ix_2$, $\bar x_{||}=x_1-ix_2$, and in
the transverse non-compact directions $x_\perp=x_0+ix_3$, $\bar
x_\perp=x_0-ix_3$.  Derivatives and potentials are defined as
$\partial_{x_{||}}=\h(\partial_{x_1}-i\partial_{x_2})$,
$A_{x_{||}}=\h(A_1-iA_2)$ and similarly for the other coordinates.
Inserting (\ref{gaugepotential}) into (\ref{zmode}) one can express
the components of $\Psi(x;z)$ without reference to the
$\tilde\varphi^{(p)}$.  Two components are obtained using the
$\Phi^{(2)}$ representation
\begin{equation}\label{zmode1}
\Psi_{11}=2i\sqrt{\rho}\partial_{x_{||}}\frac{
\varphi^{(2)}}{\sqrt{\rho}},~~~~~
\Psi_{21}=2\sqrt{\rho}\partial_{x_{\perp}}\frac{
\varphi^{(2)}}{\sqrt{\rho}},\end{equation}
and the remaining two components derive from the $\Phi^{(1)}$ representation
\begin{equation}\label{zmode2}
\Psi_{12}=2\sqrt{\rho}\partial_{\bar x_{\perp}}\frac{
\varphi^{(1)}}{\sqrt{\rho}},~~~~~
\Psi_{22}=2i\sqrt{\rho}\partial_{\bar x_{||}}\frac{
\varphi^{(1)}}{\sqrt{\rho}}.\end{equation}
Here we have written the zero mode components $\Psi_{\alpha p}$ where
$\alpha$ is a spinor index and $p$ an $SU(2)$ color index.  The
representation of the caloron zero mode given in
\cite{GarciaPerez:1999ux} has the same derivative structure.  The
fermionic zero mode given here has the normalisation
\begin{equation}
\int_{T^2\times R^2}~d^4x~\Psi^\dagger(x;z)
\Psi(x;z)=4L_1L_2,
\end{equation}
and so the normalised zero mode density is
\begin{equation}
\Psi^\dagger\Psi=\frac{\rho}{L_1 L_2}
\left(
\left|\partial_{\bar x_{||}}\frac{\varphi^{(1)}}{\sqrt{\rho}}
\right|^2+
\left|\partial_{ x_{||}}\frac{\varphi^{(2)}}{\sqrt{\rho}}
\right|^2+
\left|\partial_{\bar x_{\perp}}\frac{\varphi^{(1)}}{\sqrt{\rho}}
\right|^2+
\left|\partial_{ x_{\perp}}\frac{\varphi^{(2)}}{\sqrt{\rho}}
\right|^2\right).
\end{equation}
We have written $A_\mu(x)$, $\psi^{(p)}(z;x)$ and $\Psi(x;z)$ in terms
of the auxiliary objects $\rho(x)$, $\nu(x)$ and $\varphi^{(p)}(z;x)$.
These can be expressed in terms of contributions to the inverse of
$D^{\dagger}_x(\hat A)D_x(\hat A)$ which has the form
\cite{Ford:2003vi}
\begin{eqnarray}\nonumber 
\lefteqn{\left(D_x^{\dagger}(\hat A)D_x(\hat A)
\right)^{-1}(z,z')}\hspace{1cm}\\\displaystyle&=& 
\h(\sigma_0+i\sigma_3)e^{-\phi(z)}
K_+(z,z';x)e^{-\phi(z')}+
\h(\sigma_0-i\sigma_3)
e^{\phi(z)}
K_-(z,z';x)e^{\phi(z')}.
\end{eqnarray}
The key formulae are
\begin{equation}
\rho(x)=K_+(-\omega,-\omega;x)=K_-(\omega,\omega;x),
~~~~
\nu(x)=K_+(\omega,-\omega;x),
\end{equation}
and
\begin{equation}
\varphi^{(1)}(z;x)
=e^{\phi(z)}\frac{K_-(z,\omega;x)}{\sqrt{\rho}},~~~~~
\varphi^{(2)}(z;x)
=e^{-\phi(z)}\frac{K_+(z,-\omega;x)}{\sqrt{\rho}}.
\end{equation}
In \cite{Ford:2003vi} explicit forms for the $K_\pm$ functions were
given for the two dimensional slice $x_\perp=0$.  Although the zero
mode formulae involves $x_\perp$-derivatives they do not contribute if
$x_\perp=0$.

Using (\ref{zmode1}) and (\ref{zmode2}) the components of the smooth
non-normalisable zero modes $\Psi^I(x)$ and $\Psi^{II}(x)$ can be
recovered. For $\kappa<\frac{1}{2}$ and $z$ close to $\omega$,
$\Psi(x;z)\sim |z-\omega|^{\kappa}\Psi^I(x)$ where
\begin{equation}
\Psi_{11}^I=2ic\sqrt{\rho}\partial_{x_{||}}\frac{
\nu}{{\rho}},~~~~~
\Psi_{21}^I=2c\sqrt{\rho}\partial_{x_{\perp}}\frac{
\nu}{{\rho}},~~~~
\Psi_{12}^I=\frac{c\sqrt{\rho}}{2\pi\kappa}\partial_{\bar x_{\perp}}\frac{
1}{{\rho}},~~~~~
\Psi_{22}^I=\frac{ic\sqrt{\rho}}{2\pi\kappa}\partial_{\bar x_{||}}\frac{
1}{{\rho}},\end{equation}
where $c$ is defined by $e^{-\phi(z)}\sim c|z-\omega|^\kappa$ for $z$
close to $\omega$.  For large $|x_\perp|$ we have \cite{Ford:2003vi}
$\rho\propto |x_\perp|^{-2 \kappa}$ ($\nu$ decays exponentially)
implying that ${\Psi^I}^\dagger\Psi^I\propto |x_\perp|^{2(\kappa-1)}$
which is indeed too slowly decaying
to normalise the solution.\\[-2ex]

\noindent{\bf Acknowledgements}
 
We are grateful to DIAS, Dublin and the Institute for Theoretical
Physics, University T\"ubingen for kind hospitality. This work has
been supported by a grant from the Ministry of Science, Research and
the Arts of Baden-W\"urttemberg (Az: 24-7532.23-19-18/1)
and the DFG under contract GI328/1-2. \\[-2ex]

\end{document}